\documentstyle[12pt,epsf,psfig,epsfig]{l-aa}     % LaTeX A&A  Standard Fonts

\newcommand{\aeta}{ A\&A } 
\newcommand{\apj}{ApJ } 
 
\newcommand{\pasp}{PASP }

\newcommand{\mn}{MNRAS } 
 
\newcommand{\beq}{\begin{equation}}
\newcommand{\eeq}{\end{equation}}
\newcommand{\gam}{$\gamma$}

\newcommand{\hb}{H$_{\beta}$}
\newcommand{\ha}{H$_{\alpha}$}
\newcommand{\hg}{H$_{\gamma}$}

\begin{document}
 
   \thesaurus{06         % A&A Section 6: Form. struct. and evolut. of stars
              (02.01.2;  % Accretion, accretion discs
               08.02.3;  % (Stars: ) binaries: general
               08.09.2 BY Cam;  % (Stars: ) individual
               08.14.2;  % (Stars: ) novae, cataclysmic variables
               13.25.5)  % X--rays: stars
             }
   \title{The case of the two-period  polar BY Cam (~H0538+608~)
\thanks{Based on observations from State Research Center Special 
Astrophysical Observatory of Russian Academy of Sciences, Russia. 
}  }
\author{M. Mouchet\inst{1,2}, J.M. Bonnet-Bidaud\inst{3},  N.N. 
Somov\inst{4}, T.A. Somova\inst{4}}
 
   \offprints{M. Mouchet (mouchet@obspm.fr)}
   \institute {$^1$Observatoire de Paris, DAEC, Unit\'e associ\'ee au CNRS et 
              \`a l'Universit\'e Paris 7-Denis Diderot, F--92195 Meudon 
Cedex, France \\
              $^2$Universit\'e Denis Diderot, Place Jussieu, F--75005 Paris
              Cedex, France \\
              $^3$Service d'Astrophysique, CE Saclay, CEA/DSM/DAPNIA/SAp, 
F--91191 Gif
               sur Yvette Cedex, France  \\   
              $^4$Special Astrophysical Observatory of the Russian
Academy of Sciences, Nizhnij Arkhyz 357147, Russia}
 
\date{Received: 1996 November 26; accepted: 1997 February 3}
 
\maketitle
\markboth{The two-period polar BY Cam}{} 

\begin{abstract}
We present the results of fast temporal optical spectroscopy and photometry 
of the AM Her type system BY Cam obtained in February 1990 and March 1991. 
Emission line profiles show a complex structure and are strongly variable. 
The radial velocity studies of a sharp component detected in Balmer lines 
in March confirms the non-synchronism of the system. 
Different possible emitting regions are discussed to explain the 
characteristics of the multiple line components. 
It is shown from a computation of the line velocities and widths that the 
'horizontal' stream cannot explain the lines of intermediate widths observed 
in this system. 
A critical review of the different estimations of the periods found in BY Cam 
is presented. 
Additional periods are revealed from a re-analysis of previous optical 
polarimetric and UV spectroscopic data.
We show that because of the slight asynchronism between the white dwarf 
rotation and the orbital period, significant changes in the accretion geometry
introduce a bias in the period determination depending on the length of the 
observations. 
Large phase variations are shown to exist which are well reproduced by a 
phase-drift model in which a magnetic dipole rotates in the orbital frame 
with a period of 14.5$\pm$1.5 days. \\

%______________________________________ Do not leave a blank line here!
 
 
\end{abstract}
 
%
%________________________________________________________________
 
\section{Introduction}
Polars (AM Her type systems) are cataclysmic binaries in which a magnetic 
white dwarf accretes matter from a red dwarf star filling its Roche lobe. 
The accretion occurs along the magnetic field lines down to the magnetic poles
of the white dwarf. 
The X-rays  and  the optical polarized flux, both originating close to the 
surface of the white dwarf are modulated at the orbital period which proves
the synchronism of the white dwarf rotation with the orbital motion. 
A comprehensive review of these objects has been done by Cropper (1990).   \\
In the case of BY Cam, the situation concerning the different periods in the 
system is not clear. The X-ray source H0538+608 (BY Cam) has been identified 
as a polar by Remillard et al.(1986), on the basis of the detection of a 
circularly polarized optical flux.
This source shows two brightness states (Szkody et al. 1990), such a 
behaviour being shared by most polars. However this object is atypical by 
several aspects. 
In the UV it reveals an abnormal emission line spectrum with an enhanced NV
line and a weak CIV line (Bonnet-Bidaud and Mouchet 1987). 
This could be linked to the chemical composition of a secondary whose outer 
layers have been lost during the evolution of the system (Mouchet et al. 
1991). This is also reminiscent of what is found in some novae, suggesting a 
possible unnoticed nova-like event (Bonnet-Bidaud and Mouchet 1987). 
However no nova outburst is recorded in archive plates (Silber et al. 1992 
(hereafter SBIOR)). Noteworthily, X-ray spectra of BY Cam obtained 
with the BBXRT experiment revealed an oxygen absorption edge near 0.6 keV,
which intensity either requires  an overabundance of oxygen or  partially 
ionized material (Kallman et al. 1993).  \\

\begin{table*}
%\begin{center}
\caption{Log of the photometric and  spectroscopic observations } 
\begin{flushleft}
%\small
\begin{tabular}{ccccccccccccc}
& & &  & & & & & && & &\\ \hline
\multicolumn{5}{c}{Photometry}&&&&   \multicolumn{5}{c}{Spectroscopy} \\
date & start & end & phase  & total exp.& V mag.&&&&  start & end & 
phase  & total exp. \\ 
 & (UT) & (UT) &start-end & &&&&&(UT) &(UT) &start-end  & \\ \hline
& & &  & &  & & & & & \\ 
25 Feb. 90&17:40&20:46&0.22-1.15&3h06m&14.6      &&&& 17:39& 19:04& 0.22-0.61&       
 \\
        &     &     &         & &              &&&& 19:51& 20:41& 0.90-1.11&2h15m  
\\
10 March 91&16:44&21:10&0.67-2.01&3h26m&          &&&& 16:32& 17:42& 0.63-0.93&      
\\
        &     &     &         & &15-15.6       &&&& 17:49& 18:49& 1.02-1.27&     
\\ 
        &     &     &         & &              &&&& 18:57& 20:01& 
1.36-1.62&4h14m\\
        &     &     &         & &              &&&& 20:05& 21:06& 1.71-1.96&     
\\
12 March 91&16:43&20:13&0.12-1.17&3h30m&14.2-14.9 &&&& 16:39& 17:39& 0.11-0.32&     
\\
        &     &     &         & &              &&&& 17:44& 18:16& 0.44-0.53&     
\\
        &     &     &         & &              &&&& 18:21& 19:33& 0.63-0.92&3h29m  
 \\
        &     &     &         & &              &&&& 19:38& 20:23& 1.01-1.18&       
\\
\hline
%& & &  & & & & &\\ \hline
\end{tabular}
\end{flushleft}
%\end{center}
\end{table*}

Contemporaneous X-ray and optical observations  (Ishida et al. 1991, SBIOR)  
have revealed the presence of two very close periods around 3.3h, suggesting 
the asynchronism of the system (SBIOR). 
This was the second polar sharing this peculiarity: the first one being 
V1500 Cyg, a nova which is expected to become synchronous again in less than 
200 years (Schmidt and Stockman 1991, Katz 1991). 
A third one, RX J1940.1-1025, has been added recently, but contrary to both 
previous ones, the spin period is slightly longer than the  orbital period 
(Friedrich et al. 1996). 
From a compilation of previous optical polarimetric and photometric 
measurements of BY Cam, combined with new observations, Piirola et al. (1994) 
obtained a new determination of the shortest period (3.3308h) which 
differs significantly from the original determination (Mason et al. 1989, 
hereafter MLS).    
The presence of two close periods in this system led to search for  
 a longer period of the order of fourteen days which would be the beat 
period. 
A large set of photometric data collected over a period of 66 days, seem to 
reveal a period of seven days (Silber 1995), but a beat period of 14 days has 
also been suggested by Mason et al. (1995a,b). \\

Recently, time-resolved UV spectroscopy revealed a modulation of the line 
fluxes and of the radial velocities with the longest period identified as 
the orbital one (Zucker et al. 1995), implying an origin far from the 
accretion column, contrary to the common idea that the high ionization
UV resonance lines are formed close to the white dwarf. 
In the optical emission lines are  very complex and can show up to four
components (MLS). \\

In this paper we present an analysis of high temporal resolution spectroscopic
and photometric optical data obtained at three different dates. Partial 
results have been presented by Bonnet-Bidaud et al. (1992).
The optical emission lines, composed of components arising from different 
regions, are good tools to constrain the different periods 
present in this system and to trace the geometry of the accreting flow. 
The optical results are compared to the UV spectroscopic results. 
A critical review of the different period determinations is also presented 
in Section 5 and their interpretation is discussed in Section 6 in terms of
phase-drifts introduced by an asynchronous rotating magnetosphere. \\

\section{Observations}

\begin{figure}
\epsfxsize=8.8cm
\epsfbox[90 45 420 630]{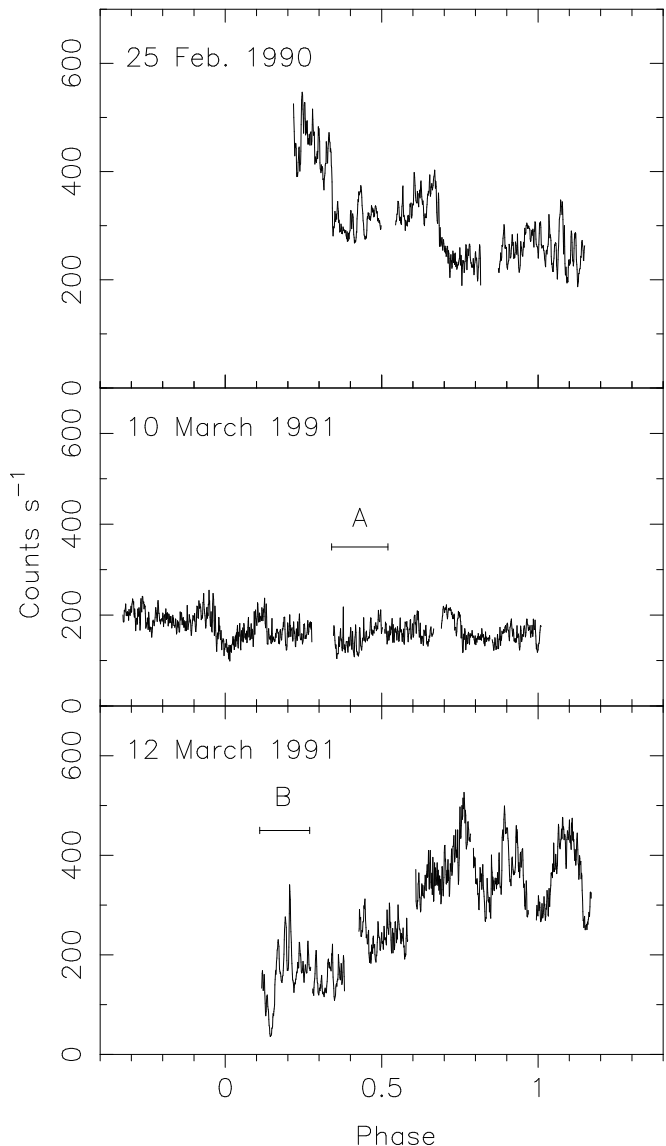}
\caption[ ]{The airmass and sky-corrected optical B light curves of BY Cam 
for the three observing dates.
The phase is computed with the trial ephemeris of Mason et al. (1989). 
The regions marked A and B are enlarged in Figure 2. 
Note the different brightness levels and the strong variability on Feb. 25 
1990 and March 12 1991.}
\end{figure}

\begin{figure}
\epsfxsize=8.8cm
\epsfbox[25 35 450 640]{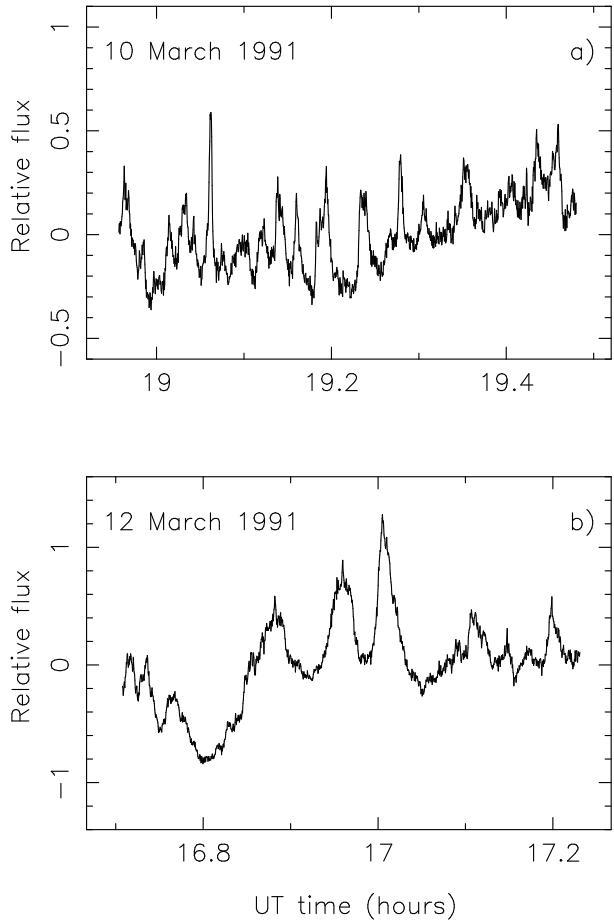}
\caption[ ]{Enlargements of the optical B light curves of BY Cam 
on March 10 (a) and on March 12 (b) for an illustration of the strong 
variability.
The plotted parts are marked A and B in Fig.1. The temporal resolution is 
$1.6~$s for both curves. The vertical scale is the deviation in \% from 
the mean flux value of the corresponding part of the light curve. }
\end{figure}

Spectroscopic and photometric data were obtained simultaneously at the AS SAO 
6m telescope in Zelenchuk (Russia) in February 1990 and March 1991. 
The observations were performed using the SP-124 spectrograph and the NEF 
photometer at the Nasmyth secondary focus of the 6 meter Bolshoi Azimuthal  
Telescope (Ioannisiani et al. 1982, Vikuliev et al. 1991). 
The spectrograph was equipped with a 1200 lines/mm grating. 
A television scanner with two lines of 1024 channels was used to record 
the sky and source spectra simultaneously in a photon-counting mode 
(see Somova et al. (1982), Drabek et al. 1986, Afanasiev et al. 1991, 
for a detailed description of the instrumentation).
The intrinsic temporal resolution is 0.033~s.
The wavelength range was 3900-4920\AA, 4030-5050\AA{} and 3950-4970\AA{} 
respectively  in Feb. 1990, on March 10 and March 12 1991. 
A wavelength calibration has been applied using a He-Ar-Ne lamp.
The aperture is a circular slit which diameter was chosen between 2 and 3",
according to the seeing. 
The corresponding spectral resolution is $\sim$2.5\AA. 
The sky spectrum has been subtracted but no flux calibration has been applied.
The log of the observations is reported in Table 1.    
Photometric data were obtained simultaneously with the spectroscopic data by 
splitting the light beam, with about 50\% of the light being sent into a 
12" aperture one-channel fast photometer. 
Continuous Johnson B data were recorded with a 0.1s resolution. 
UBVR were also acquired in one or two occasions during each observation 
and have been used to compute the optical magnitude of the 
source using star F from 3C147 field (Neizvestny 1995) as a reference star. 
The V magnitude values, reported in Table 1, show the source to 
be in a high state during the three observations, distinct from  the 
low state observed by Szkody et al. (1990) in January 1989.
When two measurements have been done during the same night,
their different values are consistent
with the amplitude of the light curve (Remillard et al. 1986, SBIOR).    \\
Throughout this paper, heliocentric phases refer to the ephemeris given 
by MLS that is used as a comparison ephemeris.
Given the uncertainty in the determination of the different periods of 
the system, the accuracy given in this ephemeris is purely formal and the 
period (3.322171$\pm 0.000017$h) is used here only as a trial period to 
facilitate the comparison with other works (see par. 5.1). \\

\section{Photometry}
 
In Figure 1 we show the B light curves obtained simultaneously with the 
spectroscopic data, at a temporal resolution of 12.8$~$s for the three 
observing dates. 
They reveal two levels of activity. On February 25 1990 and on March 12 1991 
the  large amplitude of the modulation is reminiscent of the flaring states 
observed in the optical as well as in the X-rays by Ishida et al. (1991) 
and SBIOR, while on March 10, the light curve shows a lower intensity which 
indicates a level closer to a pulsing state (Ishida et al. 1991, SBIOR). 
For both February 25  and March 12 observations, the lowest levels of the 
light curves which occur at the end and at the beginning of the observations 
respectively, are consistent with the level detected on March 10. \\ 
The higher fluxes in 25 Feb 1990 and 12 March 1991 compared to March 10 
suggest an additional source of light. 
During a flaring state, SBIOR have reported a one-peaked periodic 
light curve, which exhibits an increase by a factor $\sim$2 between the 
minimum and the maximum. The Feb. and March 12 curves also exhibit a total
variation by a factor 2-2.5, but their strong variability excludes a regular 
quasi-sinusoidal shape. 
However, Silber  (1995) reports flaring state light curves which also depart 
from a regular shape. \\ 
The March 10 light curve varies in amplitude by a lower factor of 1.5, quite 
consistent with the pulsing state light curves shown in SBIOR. 
There is no clear indication of the presence of two maxima, while these bumps 
are usually seen in the low-level optical curves with possible unequal 
intensities and separated by a variable phase extension (SBIOR, Silber 1995).\\

Apart from the large amplitude main modulation, these optical light curves
show shorter timescale variability.  
We note a significant ($\sim$ 5 min.) dip seen near phase 0.1 on March 12.
In Figure 2, are reported two blow-ups of March 10 and March 12 light
curves exhibiting  strong flaring on a typical timescale of 1-2 minutes on 
March 10 and of 3-5 minutes on March 12. 
The FFT power spectrum analysis of the data does not show significant 
power excess at these periods indicating that these oscillations are not
coherent. On March 12, quasi-periodic oscillations of the order of 
30 minutes are visible in the second part of the light curve (Fig.1), 
similar to what reported previously by SBIOR. \\

\section{Spectroscopy}
\subsection{Description of the line profiles}
Spectra averaged over the full observation have been produced for the three 
dates. 
They are similar to previously published spectra, exhibiting the usual strong 
Balmer and helium lines (Remillard et al. 1986, MLS).
In order to study the orbital and rotational modulation of the line profiles,
the spectra have been co-added with a 500s resolution which is the best 
compromise between a high temporal resolution and a good signal-to-noise 
ratio. This results in 16, 29 and 22 spectra respectively on 1990 Feb. 25, 
1991 March 10 and March 12.
The spectra reveal the presence of several variable components in both Balmer 
and He lines. 
In Fig. 3 the HeII and H$_{\beta}$ profiles normalized to the continuum are 
shown for the three epochs in order of increasing phase from bottom to top. 
Large variability with phase at a given epoch as well as strong changes 
between the three observations are clearly seen. 
Remarkably, the  HeII profile differs from the \hb{} profile in most spectra. 
The most complex profiles appear in Feb. 90 in both Balmer and HeII line 
while in 1991, they are restricted to the Balmer components only, 
the HeII line showing a much more regular phase variation.\\

\subsection{Radial velocity results}
\begin{figure*}
\epsfbox[26 60 535 500]{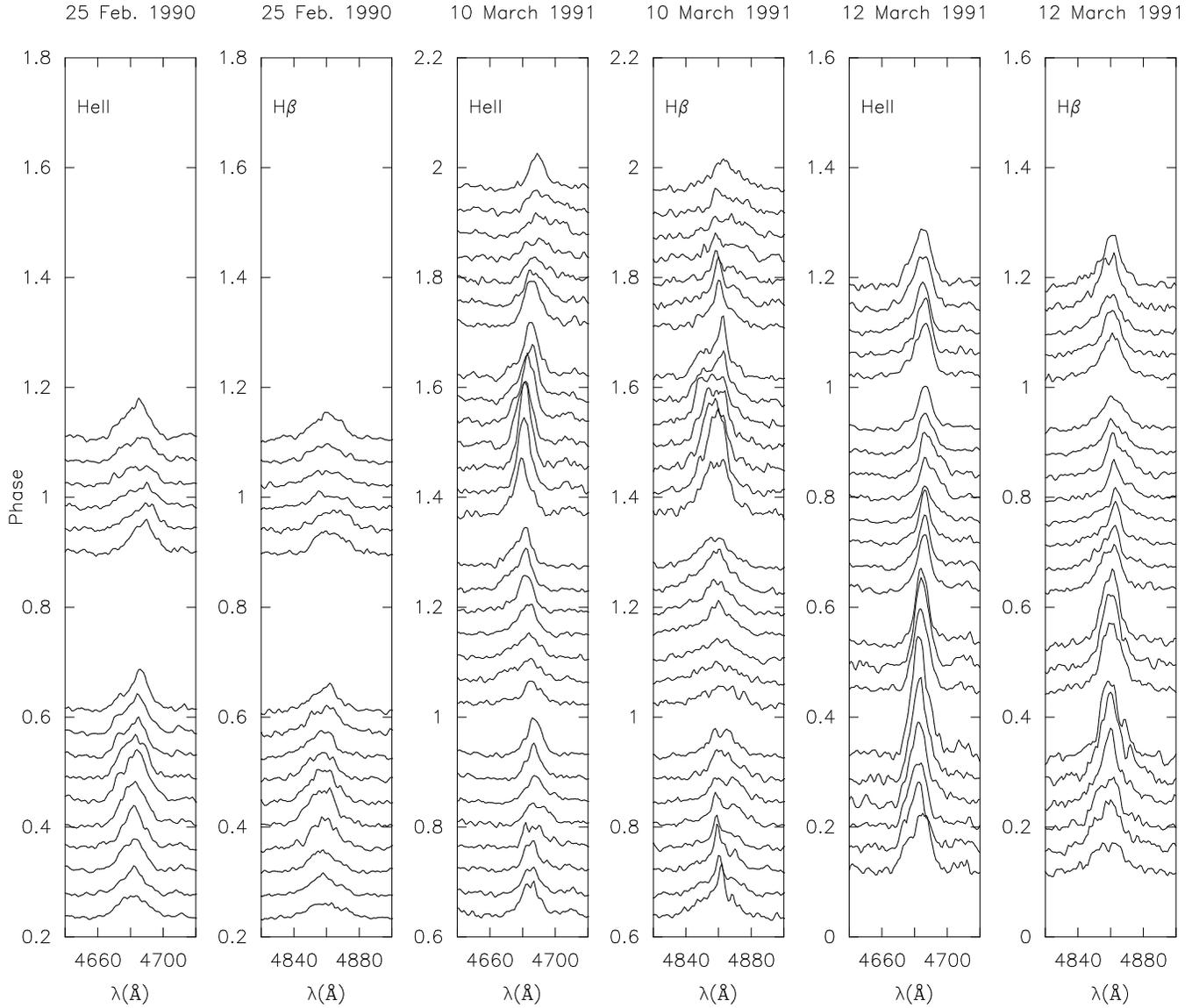}
\caption[ ]{HeII and H$_{\beta}$ normalized profiles at the same scale,
 on February 25 1990,  on March 10 1991 and on March 12 1991. 
Phase is increasing from bottom to top as shown  on the Y-axis.}
\end{figure*}

The very complex and variable line profiles observed in BY Cam make the 
radial velocity (RV) analysis very difficult. 
In addition, blends with close weaker lines might occur (f.i. HeI 4713\AA{} 
in the red wing of HeII 4686\AA) and affect the measurements. 
To isolate the different components of the complex emission lines, the 
profiles have been fitted with the sum of two or three gaussians of
variable widths, intensities and positions, using the program SPECTRE
developed by D. Pelat at Meudon Observatory. 
In this program free parameters are the width, intensity and centre of the 
gaussians as well as the adjacent continuum which is fitted with a polynomial 
of degree one. \\
 
\begin{table*}
\caption{Averaged deconvolved FWHM and radial velocity sinusoidal 
parameters of the HeII and  H$_{\beta}$ line components} 
\begin{flushleft}
\begin{tabular}{ccccccc}
 & & & & & &    \\ \hline
\multicolumn{7}{l}{  } \\
date & line & component & FWHM & $\gamma$-vel$^*$ & K amplitude  & $\phi_0$ \\ 
& & & km~s$^{-1}$ & km~s$^{-1}$  & km~s$^{-1}$ &    \\ \hline
& & & & & &    \\ 
25 Feb. 90&HeII & narrow & 200$\pm$90  & 49$\pm$16 & 265$\pm$24 & 0.608$\pm$0.013  
 \\
   &  & broad  & 1135$\pm$ 135& -50$\pm$17 & 160$\pm$24& 0.591$\pm$0.022   \\
   &  & high velocity & 405$\pm$70 & -818$\pm$15& 158$\pm$28& 0.983$\pm$0.018 \\
& & & & & & \\
10 March 91&\hb  & narrow & 165$\pm$35  & 18$\pm$35 & 230$\pm$23  & 
0.138$\pm$0.025   \\
 & HeII & intermediate width & 395$\pm$100 & -105$\pm$10 &172$\pm$14 & 
0.648$\pm$0.012   \\
  & & broad  & 1060$\pm$260  & -109$\pm$10 & 357$\pm$13 & 0.563$\pm$0.007   \\

&& & & & & \\
12 March 91&\hb  & narrow & 250$\pm$45  & -93$\pm$51  & 214$\pm$49    & 
0.389$\pm$0.0027 \\
 &HeII & broad & 1110$\pm$220   & -34$\pm$10  & 250$\pm$16 & 0.449$\pm$0.009 \\
& & & & & & \\ 
\hline \\  
\end{tabular}
\end{flushleft}
$^*$ These values should be regarded with caution due to the large uncertainty
of the absolute wavelength calibration
\end{table*}

\begin{figure*}
\begin{center}
\epsfxsize=15.cm
\epsfbox[25 48 583 571]{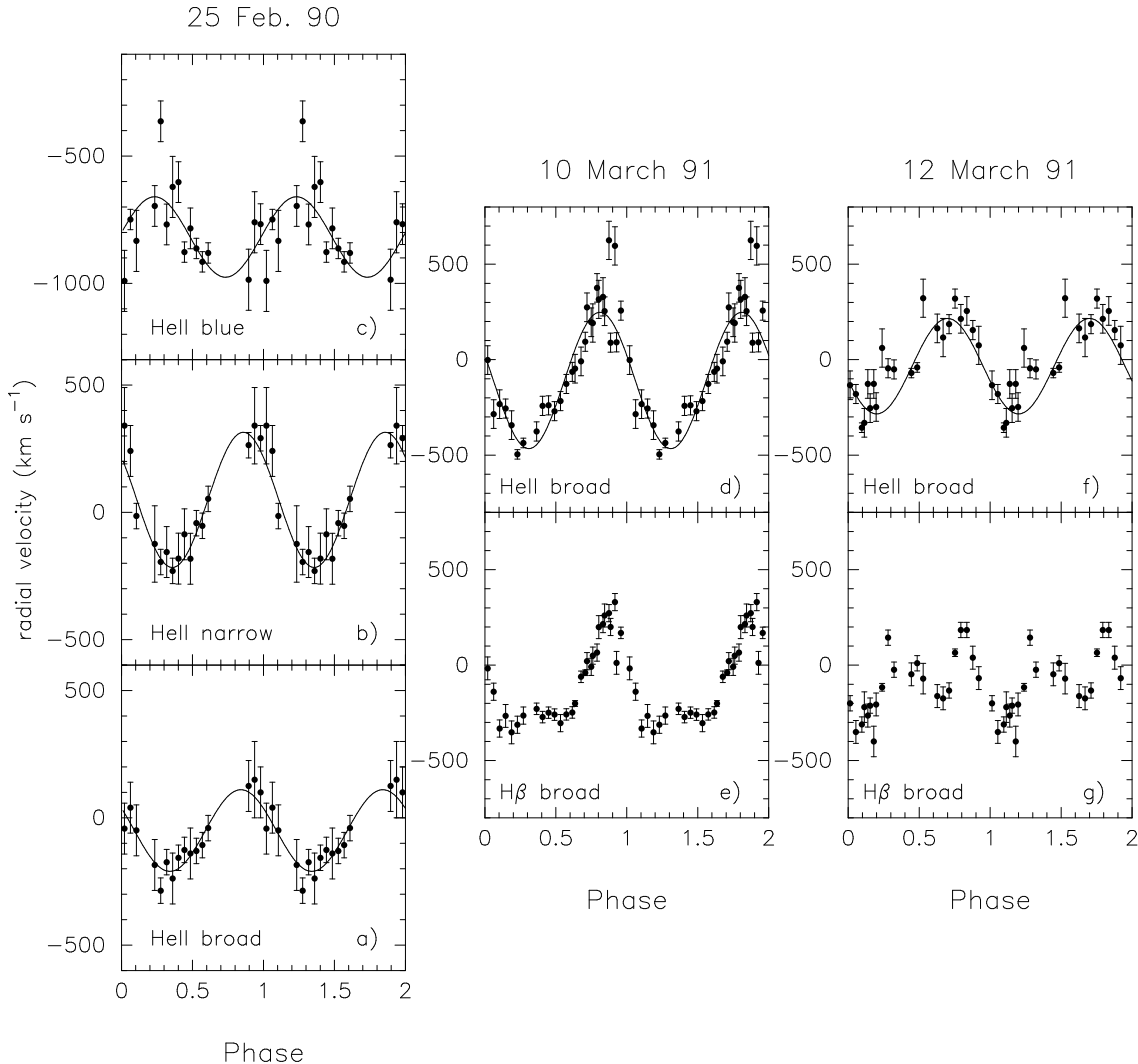}
\caption[ ]{Radial velocities against phase computed using Mason's ephemeris 
for the different components detected in the line profiles.
Left: for the three components detected in the HeII profile on 
February 25 1990: 
a) broad component; b) narrow component; c) blue high velocity component.
Middle: for the broad component of the HeII (d) and \hb{} (e) profiles on 
March 10  
Right: for the broad component of HeII (f) and \hb{} (g) on March 12. 
Best sinusoidal fits are reported (see parameters in Table 2), 
except for f and g which have not been fitted since they markedly depart 
from a sinusoidal curve (see text).}
\end{center}
\end{figure*}

The results of the gaussian fits of the HeII 4686\AA{} and \hb{} profiles 
have been used to derive the velocity curves of the different components. 
The measured radial velocities for HeII  and for \hb{} were fitted with a 
sinusoid $V(\phi)~=~\gamma+K~sin2\pi(\phi-\phi_0)$, 
$\phi$ being the phase computed from the MLS ephemeris.
The best fit parameters are reported in Table 2 where the error bars 
correspond to a $1\sigma$ deviation. 
The uncertainties given in Table 2 do not take into account possible 
systematic uncertainties in the absolute calibration. The errors 
in the absolute $\gamma$-velocity are therefore underestimated. \\
Multiple components are clearly present  and the deconvolution may not 
be unique in some cases. The FWHM of the different components are given
in Table 2, corrected for the instrumental response.
In February 90, the HeII line is wide and exhibits a complex profile: in 
addition to a narrow width (FWHM$\sim$200~km~s$^{-1}$) component 
and a broad (FWHM$\sim$1140~km~s$^{-1}$) one, a very blue high velocity 
shoulder is detectable  mostly between phases 0.4 and 0.6. 
The RV measurements of the narrow and broad  components can be satisfactorily 
fitted with a sinusoidal curve, while the measurements of the blue high 
velocity component are more scattered (Fig. 4a,b,c). 
The narrow and broad components appear in phase but with different amplitudes. 
The high velocity component lags the two others by 0.4 in phase.
It appears not to be similar to the very high velocity component detected by 
MLS. Indeed there is no clear indication of a red counterpart at opposite 
phases and the observed modulation is not compatible with an amplitude as 
large as $\sim$ 800 km~s$^{-1}$ found by MLS. 
Its low amplitude and its large negative \gam-velocity seem to favour an origin
close to the white dwarf, such as a Zeeman component which  has been 
already proposed by McCarthy et al. (1986) to explain the stationary high 
velocity narrow component found in the HeI lines of QQ Vul. 
At the same epoch, the Balmer lines show more complex profiles than the HeII 
(Fig. 3) and sinusoidal fits cannot be quoted unambiguously.  \\
In March 1991, the Balmer and HeII lines are strongly variable. 
The HeII line essentially exhibits two components, one broad 
and one of intermediate width (FWHM$\sim$400~km~s$^{-1}$) with no evidence of 
the narrower (FWHM$\leq$ 200~km~s$^{-1}$) component seen in Feb. 90. 
On March 10, radial velocities of the intermediate width HeII component are 
satisfactorily described with a sinusoidal fit, while on March 12, this 
component is definitely present but its RV curve departs from a sinusoidal 
curve, being quasi-stationary between phase 0.6-1.05.  
For both dates, the radial velocities of the broad component in HeII have 
been fitted, though on March 12 they show a significant distortion around 
phases 0.2-0.4 (see Fig. 4f).
The resulting amplitude is lower by 25\% on March 12 than on March 10 and 
the radial velocities on March 10 lag the ones on March 12 by 0.11 in phase 
(Fig. 4d,f).\\
The RV measurements of the broad component in \hb{} also strongly depart 
from a well-defined sinusoidal curve, particularly on March 12 (Fig. 4e,g).  
The distortions present in these RV curves are characterized by a shoulder  
present at phases 0.2-0.5, at the two dates, being markedly pronounced on 
March 12. It is reminiscent of what has been seen in the flaring state data 
presented by SBIOR.
The RV curves cannot be described by a sinusoidal curve in this case.
It may indicate either the presence of an additional
redder component or a partial eclipse of the broad component at these phases.  
Noteworthily, similar deviations have been reported for V1500 Cyg by Kaluzny 
and Chlebowski (1988) who have attributed them to the contribution of
a second accreting column. \\

At the difference of the HeII line, the \hb{} and \hg{} profiles show the
presence of a narrow (FWHM$\sim$150-250~km~s$^{-1}$)  component. 
The peak is clearly observed around phases 0.5 to 0.9 and repeatable from 
cycle to cycle.
The velocity curve of this component is shown in Figure 5.
This component being clearly detectable at specific phases only, 
the corresponding fits are thus less constrained. 
The amplitude are similar but an obvious lag in phase is observed for the two 
observations separated by two days when data are folded with the trial period 
($3.322171$h) derived by MLS. \\
This phase difference ($0.251\pm0.037$) is consistent with an orbital period 
($1.8\pm0.3$\%) longer than the trial period. 
The best period derived from our narrow-line observations only will be 
($3.3817\pm0.0085$h). This is slightly longer, but consistent within the error 
bars, with the value derived by SBIOR from the H$_{\alpha}$ narrow line 
velocity. This provides an independent confirmation that the period 
deduced from the narrow-line measurements, usually associated with the orbital 
period, is significantly longer than the period derived from polarimetry.
We investigate more about the nature of the two periods in Section 5.\\ 

A broad component is obviously present in all data, but its 
characteristics and its corresponding sinusoidal parameters
can only be safely derived for the HeII component. Its width 
and K amplitude are typical of those measured in other polars and the 
origin of this component is attributed to regions down to the accretion column
(Ferrario et al. 1989, Mouchet 1993b).
In 1991 its relative phasing with respect to the narrow \hb{} component  
is 0.42 and 0.06 on March 10 and March 12 respectively, confirming the 
asynchronism of the system. \\

\begin{figure}
\epsfxsize=8.8cm
\epsfbox[60 120 370 375]{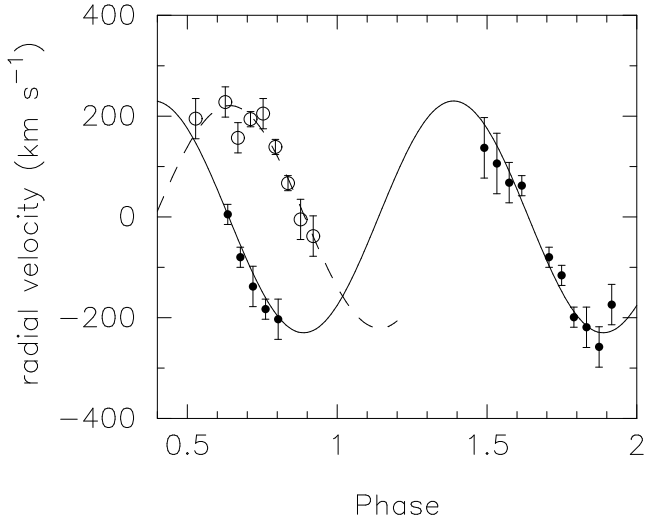}
\caption[ ]{
Radial velocities versus phase computed using Mason's ephemeris  for two 
observations separated by two days of the narrow
component of H$_{\beta}$ on March 10 (filled circles) and on 
March 12 (open circles). Data are not repeated, March 10 covering more 
than 1.3 cycle. 
For both dates, the data have been offset from the \gam-velocity 
value derived from the fit. The corresponding best sinusoidal fits are shown.
Note the significant phase lag which can be accounted for by a 
revised period of P=($3.3817\pm0.0085$h).}
\end{figure}

\section{Discussion}

\subsection{Emission lines from the horizontal stream}

The true location of the emission line regions in polars is still a subject of 
discussion.
Two main components are usually considered: a broad line associated with the 
accretion column and linked to the rotation of the white dwarf and a narrow 
line associated with the heated face of the secondary, linked to the orbital 
period.
In the case of the complex multi-component profiles observed in BY Cam, the 
situation is more intricate and several other regions may contribute to 
build the resulting profile.
In an effort to reproduce the non-standard observed components, we have 
investigated a region which has not been yet considered in detail, i.e. the 
horizontal stream of matter that links the secondary to the accretion capture 
region, roughly defined by the so-called magnetic capture radius. \\

\begin{figure}
\epsfxsize=8.8cm
\epsfbox[30 45 500 705]{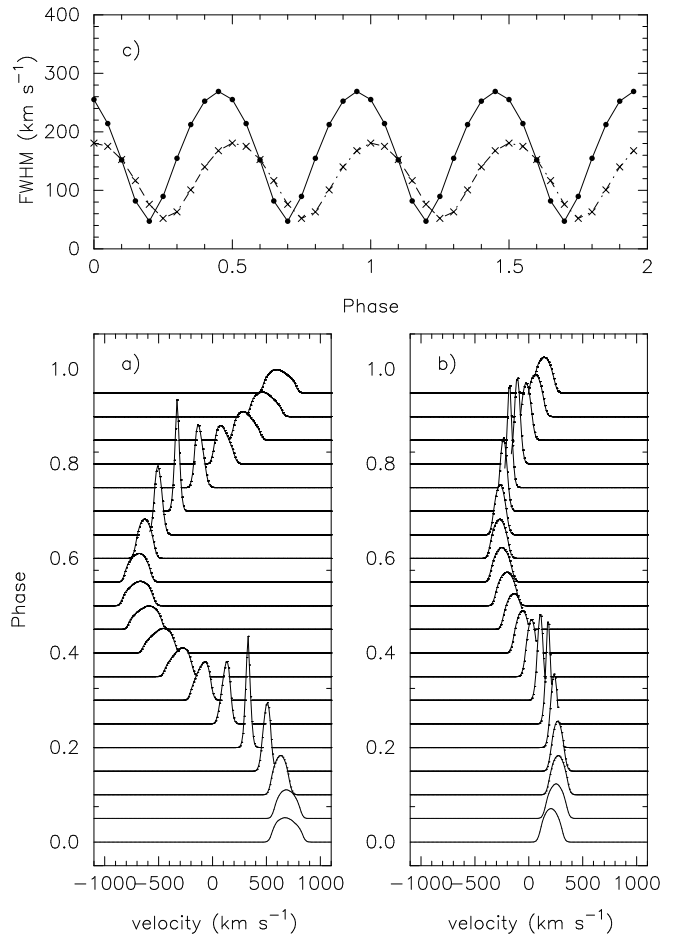}
\caption[ ]{ Simulated profiles for emission lines arising in the
'horizontal' stream, for an inclination angle of 70$^{\circ}$ and a 
white dwarf mass of  1M$_{\odot}$. A resolution of 1\AA{} has been assumed. 
Left (a): extension of the emitting region from 0.3 to 0.5 times the
projected distance between the white dwarf centre and the inner Lagrangian 
point. 
Right (b): same as (a) for a smaller extension from 0.7 to 0.9 times 
this distance. 
Upper panel (c): the FWHM versus orbital phase for the two sets of simulated 
profiles (full line: case a) ; dotted line: case b)). 
The origin of the phase is taken at the inferior conjunction of the secondary.
See text for more details.
}
\end{figure}

Before being captured by the magnetic field lines,  material forms a stream 
of matter in the orbital plane (Liebert \& Stockman 1985) (referred below 
as the horizontal stream, as opposed to the accretion column which refers to 
the out-of-plane part of the accreting flow).  
This stream has been suggested as a possible contributor to the emission line 
profiles (Mukai 1988, Mouchet 1993a). The profile characteristics of a 
component formed in the horizontal stream mainly depend on the position and 
on the extension of the emitting region.
We have simulated the emission from the stream, assuming that its path is 
given by the ballistic trajectory as computed by Lubow \& Shu (1975). 
This stream is supposed to extend down to a distance $r_c$ 
at the capture point where the ram pressure in the horizontal stream equals 
the magnetic pressure (see more details in Bonnet-Bidaud et al. 1996).  
To evaluate  $r_c$, we have supposed that the lateral extension of the stream 
is constant, being taken equal to $10^9$ cm (Lubow \& Shu 1975, Mukai 1988). 
We assume a  1 M$_{\odot}$ white dwarf mass, a typical accretion rate
of 10$^{-16}$~g~s$^{-1}$, and a dipole geometry with 
a typical polar field of 40 MG consistent with the evaluations by 
Cropper et al.(1989) and Piirola et al.(1994). Note however that 
Mason et al.(1995b) suggest evidence of a more complex field geometry 
with  accretion onto a multipole magnetic field. 
For the line profile simulation, free parameters are the mean location of 
the emitting region, its extension and the emissivity of each part of the 
region together with the mass of the white dwarf, the inclination angle and
the orbital period.
Since no detailed physical model is available at present to give the exact 
emissivity in the lines, we made the simple assumption of an  emissivity 
through the emitting region defined as a gaussian law with a width equal 
to the extension. The emissivity is taken null outside the chosen portion of 
the stream. 
The computed phase-resolved profiles have been convolved with a spectral 
resolution of $1~$\AA{}. As illustration, Figure 6 shows different resulting 
profiles computed at twenty equally spaced orbital phases for two selected 
emitting regions, each defined by their projection onto the line of centres 
and expressed in units of the distance $X_L$ between the white dwarf and the 
inner Lagrangian point. 
Fig. 6a corresponds to a stream extension between  0.3 to 0.5 times $X_L$, 
and Fig. 6b to a region far from the white dwarf, between 0.7 and 0.9 times 
this distance.  
An inclination angle of 70$^{\circ}$ has been assumed. 
These profiles clearly show a two-peaked orbitally modulated width (Fig. 6c).
We stress that the lines produced in the stream would therefore in general be
strongly variable in shape and width, appearing alternatively narrow and 
broad within the same cycle. Such lines would thus be difficult to extract 
by the standard two-component analysis.
However such high temporal and spectral resolution measurements are difficult 
to derive from most data. 
Therefore only the phase-averaged full widths at half maximum  expected from 
different zones of the stream  have  been compared to the data. 
In Figure 7 these intrinsic widths are reported against the amplitudes K of 
the RV curves of the lines emitted by  seven regions of variable positions 
(a) and extensions (b). The inclination angle increases along the curves 
from 10$^{\circ}$ to 80$^{\circ}$.
As expected, the wider the emitting region is, the larger the width is.
Maximum values of the order of FWHM $\sim$ 400 km~s$^{-1}$, as observed, 
can only be reached for a contribution all along the stream. 
High K values are obtained for the parts of the stream which are the 
closest to the white dwarf. Lines formed in the stream at a projected distance 
from the white dwarf reasonably larger than $0.1~ X_L$, the value which 
corresponds to a typical capture radius at $\sim$21~R$_{wd}$, have RV 
amplitudes restricted to values lower than  $\sim$ 800 ~km~s$^{-1}$ for a 
reasonable (i$\leq 70^{\circ}$) inclination of a non-eclipsing system. \\ 
For higher white dwarf masses (1.4 M$_{\odot}$) the curves  are shifted up,  
of the order of 10\%, towards greater RV amplitudes and widths. 
The plotted curves are computed for a system of a 3.33h orbital period, 
but the values do not increase by more than 20\% for a typical shorter
period of 2h.  \\

\begin{figure*}
\begin{center}
\epsfbox[40 54 560 365]{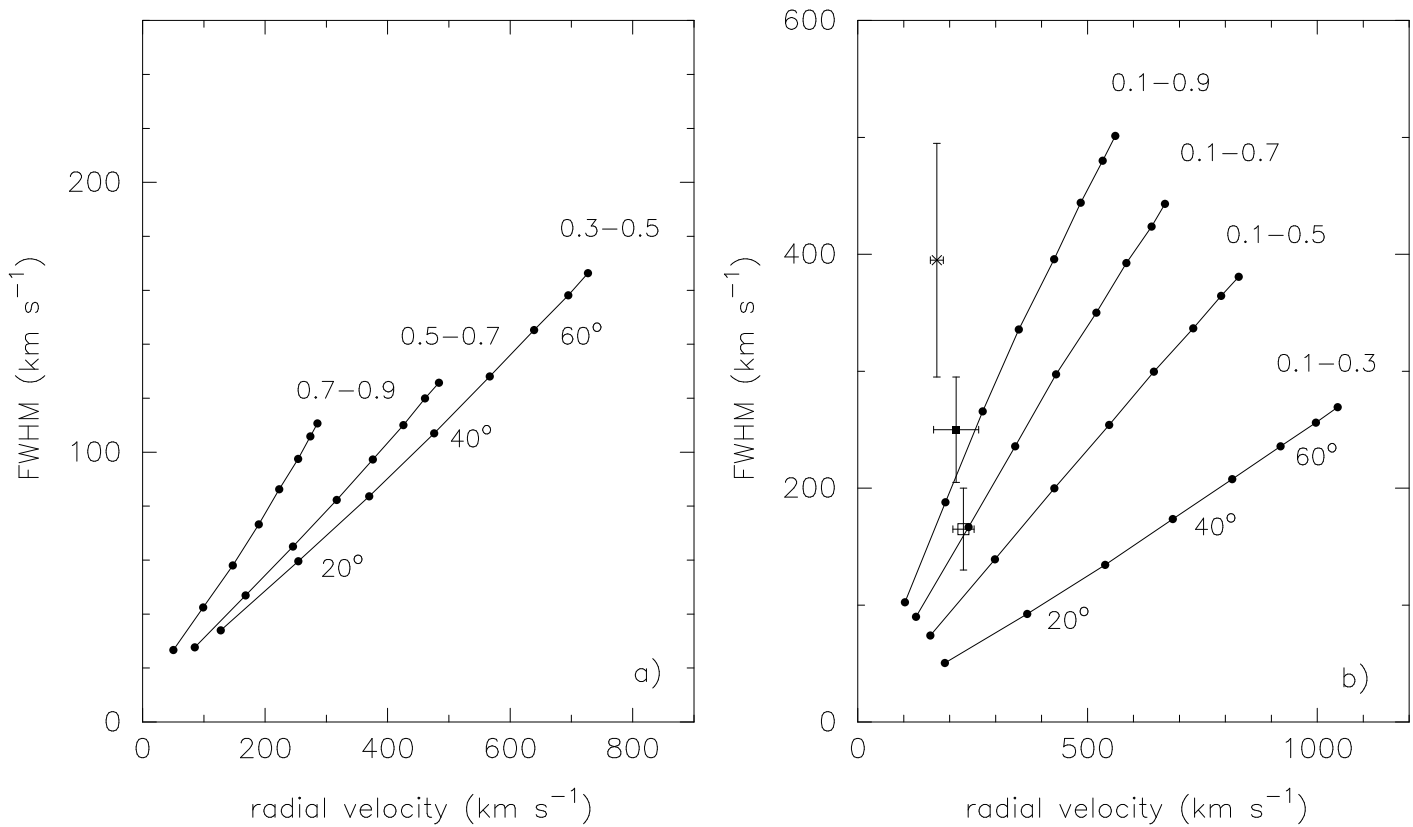}
\caption[ ]{
The orbitally averaged intrinsic  FWHM 
(in km~s$^{-1}$) versus the amplitude of the radial velocity curves  
for emission lines formed in the 'horizontal stream' for different values 
of the inclination angle (increasing from 10 to 80$^{\circ}$ along the curves)
for two series: at left (a) for a given extension at different positions in 
the stream; 
at right (b), for an extension increasing from a region close to the white 
dwarf.
The curves are labeled with the corresponding projected extensions of 
the emitting region, expressed in units of X$_{L}$,the distance between the 
white dwarf centre and the inner Lagrangian point.
The measurements corresponding to the \hb{} narrow component on 
March 10 (empty square) and 12 (full square) and to the HeII intermediate
 width  component on March 10 (cross) are also reported.}
\end{center}
\end{figure*}

These stream properties can be compared with the characteristics of
the components that were identified in BY Cam. 
Narrow components have been found in Balmer lines as well as in the HeII line 
at different epochs (see Table 2). 
The properties of the narrow component measured in Feb. 1990 spectra, can be 
satisfactorily ascribed to the stream but its width value is not much 
constrained.  
A large region starting close to the white dwarf and a low inclination angle 
(i $< 30^{\circ}$) can account for the measured values.
On March 10 1991, the width and the amplitude of the \hb{} narrow line are 
compatible with an origin in the stream, for an inclination angle lower 
than 30$^{\circ}$ and a large extension of at least 0.1-0.5X$_L$ (Fig.7). 
The data on March 12 are marginally accounted for by  an extreme case where 
the complete stream emits. 
Alternatively, the fact that this narrow component is mainly visible at 
phases 0.6-1.0 when its radial velocity passes from red-to-blue may favour 
instead an origin in the heated face of the secondary (see below).  \\ 
More interesting is the intermediate width component measured in the HeII 
line on March 10 1991. Such width is not easily produced on the secondary.
It can be produced in the stream for extended regions and high 
inclinations, but would have rather high velocity amplitudes.
The HeII intermediate width component cannot satisfy both requirements 
(K amplitude and FWHM) for an origin in the stream, the amplitude being
much lower than expected from the stream (Fig.7). 
In addition, its relative phasing with the narrow peak seen in \hb{} (0.51)
is far from the expected value if this last component is indeed formed in the 
hemisphere of the secondary. 
In fact, the simulations show that the blue-to-red phasing of the stream
 component related to the position of the secondary is constrained to an 
interval of phase of 0.75-0.85 depending on the exact position, where 
phase zero refers to the inferior conjunction. \\

On the basis of these simulations of the line emission in the stream, we
can also  reconsider the possibility that the UV lines arise from such
a region, as it has been suggested by Zucker et al. (1995).
The mean K radial velocity amplitude of 370 km~s$^{-1}$ for CIV and of 245 
km~s$^{-1}$ for HeII can be easily reproduced with an  origin in the stream. 
The evaluation of the UV line widths is made more difficult in the case of 
the wide NV, CIV and SiIV doublets which are not resolved with IUE. 
However, in the case of the HeII 1640~\AA{} line, the blend is made 
of two very close lines and does not affect the determination.
Using archive spectra, the average measured FWHM of the HeII line is 
of  the order of 8~\AA.
The IUE profiles along the dispersion direction being well represented by 
gaussian profiles of about 5.5$\pm0.3$\AA{ } FWHM at this wavelength 
(Cassatella et al. 1985). After deconvolution the intrinsic 
width of the UV HeII line is then of $\sim 1060\pm50~$km~s$^{-1}$.
The long exposure time may be responsible for part of the broadening. However
such a high value is incompatible with the values expected from a stream 
emission (see Fig. 7), unless the stream is excessively elongated.
We note that due to the de-synchronization of the system, the position of 
the capture radius is in fact modulated at the beat period between the spin 
and the orbital periods (see Section 6).
We have computed this position for given co-latitude angles $\theta_d$ of the 
magnetic field and for different values of the longitudinal angle $\psi_d$,
varying from 0 to 360$^{\circ}$, to mimic the variable configuration of the 
magnetic field.
We find that the capture distance r$_c$ varies very little with $\psi_d$,  the
strongest modulation (from $\sim$21 to 27 R$_{wd}$) being found for a high
inclined magnetic axis ($\theta_d=85^{\circ}$). 
We obtain a maximum elongated stream for a minimum magnetospheric radius of  
$\sim20.6~$R$_{wd}$, which corresponds to a projected distance on the line of 
the centres of $\sim$0.1 X$_L$. 
Even an emission from a stream extended down to this value cannot account 
for the widths of the lines.
An alternative origin of the UV lines in the accretion column  is proposed 
in Section 6. \\ 

\subsection{The heated hemisphere of the secondary} 

As suggested above, the narrow component seen in the \hb{} profile in March 
1991 might be thought to arise from the X-ray heated hemisphere of the 
secondary.
A lower limit of the white dwarf mass can thus be derived using the radial
velocity amplitude of this narrow component.
In the case of the extreme hypothesis that the K value represents the motion 
of the centre of the red dwarf, values of the amplitude
K between 200 and 250 km~s$^{-1}$ can be obtained for a white dwarf mass$>$ 0.6
M$_{\odot}$ and an inclination angle i larger than 40$^{\circ}$. 
If,  to take into account the decentered position of the gravity centre of 
the illuminated hemisphere, we apply a K-correction of 1.3, typical value 
derived from previous studies (Mouchet 1993a), the permitted range of WD 
masses is then 0.8-1.4 M$_{\odot}$, in agreement with the lower limit found 
by Ishida et al. (1991), and i should be greater than 50$^{\circ}$. \\
In contrast, for the UV lines, these large widths exclude a dominant origin 
from the heated secondary. 
The heated hemisphere has been suggested by Zucker et al. (1995)
to contribute some fraction of the NV line. Indeed a narrow component has been
clearly detected in the UV lines of AM Her using high resolution spectra 
(Raymond et al. 1995).  
It has also been invoked to contribute to the UV HeII line of V1500 Cyg 
(Schmidt, Liebert \& Stockman 1995), but the formation of this line in 
V1500 Cyg is favoured by the presence of a high temperature white dwarf 
which strongly contributes to the illumination of the secondary.\\
 
\subsection{Ephemeris: orbital and spin periods}

\begin{figure}
\epsfig{file=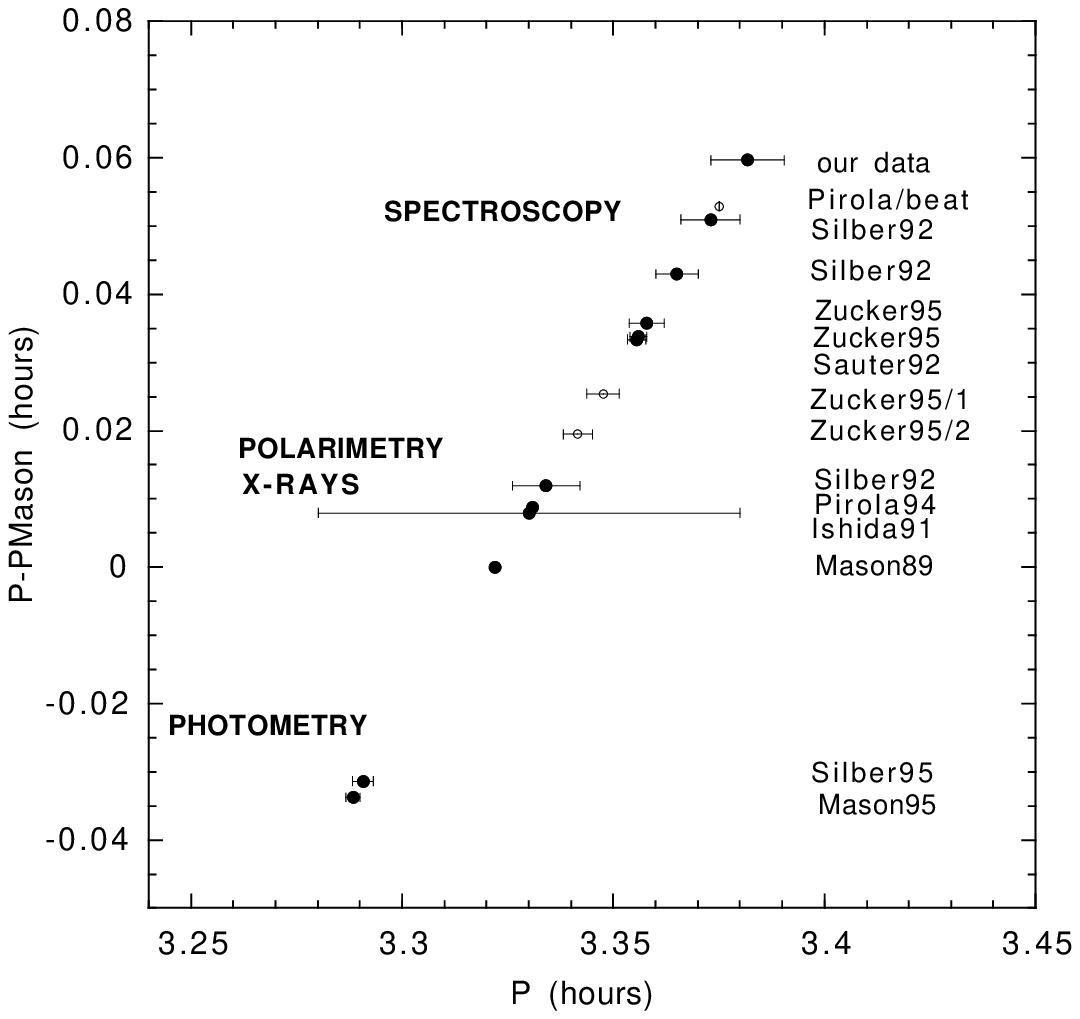,clip=,width=8.8cm,bbllx=110,bblly=400,bburx=440,bbury=710}
\caption[ ]{
Schematic representation of the different periods found in BY Cam with the 
identification of the origin of the data. 
The open symbols are the different periods determined in this paper.
Note the distribution of points according to the nature of data and the
nearly continuous distribution.
}
\end{figure}

\begin{figure}
\epsfxsize=8.8cm
\epsfbox[90 80 430 615]{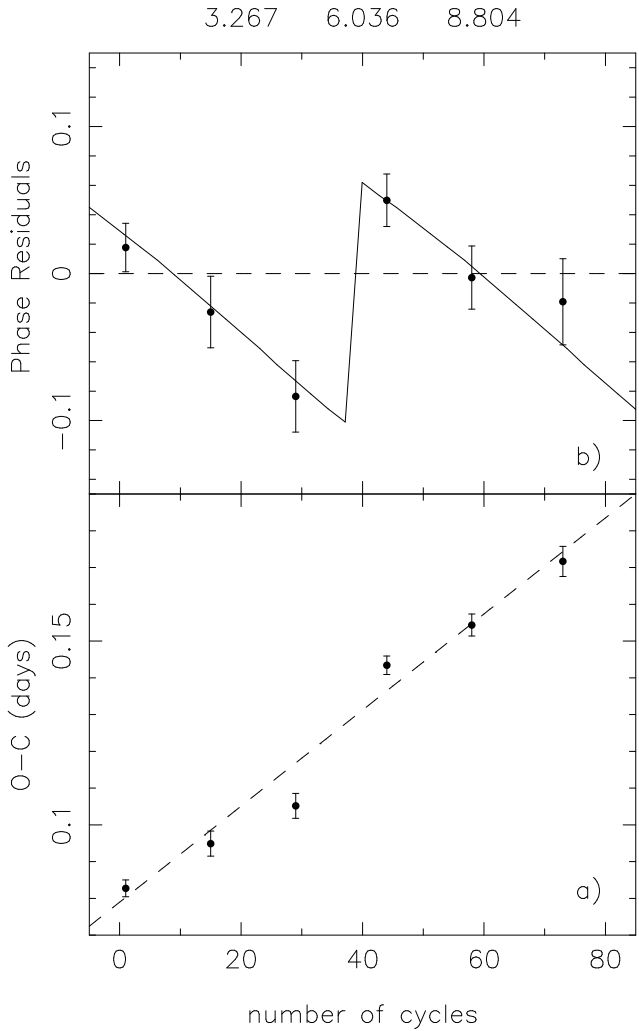}
\caption[ ]{
a): O-C values of the time arrivals  for the radial velocity measurements 
of the UV emission lines given by Zucker et al. (1995). Dotted line is 
best linear fit. b): Phase residual values after subtraction of the
best fit. Note the jump between the third and fourth points. The dot-dashed
line represents the prediction from a phase-drift model for a magnetic dipole 
co-latitude of 12$^{\circ}$ and a beat period of 14 days (see text). 
}
\end{figure}

The determination of an exact ephemeris of BY Cam appears to be a tedious 
task because of the difficulty in defining adequate stable time markers
in an asynchronous binary system.
Mason et al. (1989) first determined a high accuracy ephemeris based on a 
recurrent sharp drop in the circular polarization data that was assumed to 
track correctly the white dwarf rotation. They noted however that their
resulting ephemeris was only of apparent high accuracy, given the 
highly irregular behaviour of the source and the large gaps in their
data set. Strictly speaking, this first tentative ephemeris with a period of 
(3.322171$\pm 1.4~10^{-5}$h)
can therefore only be used as a comparison ephemeris.
A possible alias period near 3.351~h was noted by these authors.   \\
The existence of a different longer period (3.373$\pm0.007$h) was  uncovered 
by SBIOR from an analysis of the H$_{\alpha}$ narrow line velocity. 
A similar value was also derived from the line intensity variations. 
The narrow line emission is supposed to track the secondary motion and 
therefore to provide the value of the orbital period. 
The 1.3\% difference with  Mason's  period was then interpreted 
as an evidence of a slight degree of asynchronism of the same type as 
observed in the classical nova-polar V1500 Cygni (Stockman et al. 1988). \\
However as more data accumulate, the difference between the two periods 
becomes less clear. The existence of the longer period was confirmed 
by more data on the H$_{\alpha}$ narrow component (Sauter 1992, quoted in 
Zucker et al. 1995) and recently by the analysis of the UV emission line 
velocities (Zucker et al. 1995)  around a value (3.3558$\pm0.0020$~h), 
consistent with Sauter's orbital period. 
For the shorter period, additional polarimetric data obtained by Piirola
et al. (1994) were found to be consistent with a short value 
(3.3308196$\pm3.4~10^{-6}$h), largely outside the error quoted in the initial
high accurate ephemeris of MLS.
A possible increase of the photometric/polarimetric period has also been 
suggested by Piirola et al. and Mason et al. (1995a).
More recently, an even shorter period has been reported based on long term
 photometry  by Mason et al. (1995b) and Silber (1995) and has been 
suggested to be a combination of the spin and the orbital period.
Figure 8 shows the different periods presently reported for BY Cam.
Ranging from 3.29h to 3.38h, they demonstrate the very confusing situation 
for this source. \\

In an attempt to clarify this situation, we have performed a critical
re-analysis of existing data, giving particular attention to possible 
alias periods introduced by the large gaps between different data sets.  
These gaps cause a loss of the cycle number and prevent the usual 
(O-C) analysis from being performed. 
We used a minimization method in which an estimator is computed
from the individual timings as :\\
\begin{displaymath}
E(P) = \frac{2K}{N(N-1)}\sum_i~\sum_{j<i}\frac{[(t_j-t_i)/P]^2}{\sigma_{ij}^2}
\end{displaymath}
\begin{displaymath}
with :  K = \frac{2}{N(N-1)}\sum_i~\sum_{j<i} (\sigma_{ij}^2)
\end{displaymath}

where P is the trial period, t$_i$, t$_j$ the individual timings, 
$\sigma_{ij}$ the associated errors and N the total number of data points.
The K factor is introduced to normalize the estimator to the errors.
The best period is defined at the estimator minimum.
The relevant parameters $(t_j-t_i)/P$ are not independent variables, and thus
the estimator does not follow a $\chi^2$ distribution.
The statistical significance ({\it Prob}) was derived from the estimator 
distribution computed by a Monte Carlo method. 
We find that this method gives results comparable to the Lomb periodogram 
(Lomb 1976)  but has the advantage of allowing a weighting of the data. \\
Using this method, we first  tried to combine our optical radial 
velocity narrow line measurements with other measurements (MLS, SBIOR). 
These four collected data points span nearly five years in total.
The best period is found around 3.38~h, with a 2-day alias at 3.16~h which 
corresponds to the separation of our last two points, but a number of aliases
close to this best value obviously prevents a more accurate determination 
of the period. \\
We have also applied this method to the set of data published by Piirola et al.
(1994) which gathered  timings of the circular polarization dips and of a few 
photometric broad minima.
In the resulting periodogram, the period claimed by Piirola et al. is 
evident (3.3308$\pm$0.0004~h), but another longer period at a similar level 
of significance ({\it Prob}$~<~2~10^{-5}$) is also present at 
3.3749$\pm4~10^{-4}$~h. 
We note that this last period is consistent with the value derived by SBIOR. 
A F-test performed to compare the (O-C) residual distributions for the two 
periods gives no significant difference. 
The polarimetric data collected over ten years therefore reveal that both 
the short and long periods coexist in the source.  
Though at a less significant level, a short beat period at a value of 
3.2907$\pm$0.0004~h is also present, compatible with the value derived by 
Silber (1995) and Mason et al. (1995b) from long term photometry 
(see Section 6).\\

The discovery of a periodicity in the UV line velocity at a period
of 3.3558~h (Zucker et al. 1995), close to the refined 3.3554~h long 
period value of the narrow \ha{} velocity reported by Sauter (1992), 
was unexpected, since it suggests an origin far from the white dwarf for the 
UV lines.
We have reanalysed the data of Table 1 of Zucker et al. (1995).  
In this case, the accuracy of the data is such that one can keep track of 
the cycles through the full 11 days of the observations. 
A standard (O-C) method can therefore be used instead of a periodogram 
search and provide a more accurate measure of the phase variations inside 
the observations. 
We have fitted the NV radial velocity data of the six separate observations 
with a sine curve at Mason's period (MLS), used as a trial period. 
The phase residuals (O-C) are plotted in Fig. 9. 
The evident linear trend clearly confirms that Mason's period does not 
correctly track  the NV velocity. 
The best period determined from the trend is P=(3.3535$\pm0.0029~$h), 
in accordance with the value derived by Zucker et al. (1995). 
However after subtraction of this linear fit,
significant residual values are clearly revealed, defining a saw-tooth shape
variation through two clearly separated sets of three points. 
The mid-break corresponds to a phase shift of 0.133. 
If the two sets of points had been fitted individually, they would have
produced two different periods (respectively of 3.3415$\pm0.0035$h  and 
3.3475$\pm0.0038$h for the first and the second sets), each of them shorter 
than the mean period derived using all points.
The different "alias" periods are shown in Fig. 8, together with all the
other period determinations. The period determinations in BY Cam appear 
roughly distributed in three separate groups depending on the nature of the 
data with long values derived from spectroscopy, intermediate values from 
polarimetry and X-rays light curves and short values derived from optical 
photometry. But as illustrated by the UV data, there is an apparent continuum 
of values through the different groups (see discussion in Section 6). \\

\section{Open questions and conclusions}  
Though it has been extensively observed, the magnetic cataclysmic binary 
BY Cam  is far from been well understood. 
The debate about the periods has been reactivated by the study of the 
UV emission lines by Zucker et al. (1995).  
They demonstrate the presence of the so-called orbital period 
in the radial velocity measurements of these lines and are led to assume 
a formation of the UV lines far from the white dwarf, in the orbital 
plane.  We have shown above that, because of their broad widths,
the bulk of these lines cannot be produced either in the horizontal stream 
or in the heated hemisphere.
The most natural contribution to the UV lines is the accreting column
out of the orbital plane, as it has been previously proposed for several 
polars, based on the fact that their orbital variations are in phase with 
the broad optical components (Mukai et al. 1986, de Martino 1995). 
We show below that the temporal behaviour of the UV lines is indeed 
consistent with this origin. \\
The scattering of the periods found in BY Cam (Fig.8) can be explained by 
a scenario, in which, in the orbital frame, the position of the rotational 
axis is moving slowly with the beat period of about fourteen  days.
The accretion column is thus formed along different field lines according 
to the beat phase. 
Schematically, the accreted material is slowly dragged by the magnetic field 
during the relative motion of the white dwarf, up to a certain 
extent when the accretion is no more possible along these lines.  
Accretion has then to occur at the opposite pole, causing the jump observed in 
Fig. 9. \\
To compute this effect, we have calculated the position, on the white dwarf 
surface, of the footprints of the field lines which intercept the orbital 
plane at the capture radius. This is done for a fixed given co-latitude 
angle $\theta_d$ of the dipole magnetic field and for different values of 
the longitudinal angle $\psi_d$ which measure the different phases inside 
the beat cycle. 
When the capture region is close to the white dwarf, the accretion spot is 
significantly distinct from the magnetic pole and for different capture 
regions, its location on the white dwarf surface varies.
We find that the co-latitude $\theta_w$ of the footprint, with respect to 
the rotational axis, is weakly variable with the beat phase, while its 
longitude $\psi_w$, defined in the orbital plane with respect to the line 
joining the centres of the two stars, may strongly vary along the beat cycle.
This implies a lag of the impact spot with respect to the magnetic pole
and the resulting phase drift would be interpreted as an apparent period 
longer than the true spin period. \\
If furthermore, one assumes that the accretion occurs in the opposite 
hemisphere as soon as the threading point is situated at an angular distance 
from the magnetic axis larger than $90^{\circ}$, then at one time, 
the computed longitude abruptly goes down to low values and increases again 
in the cycle.     
We find that the  sudden change of the longitude value, due to the switching 
of the accretion, occurs before the longitude reaches 
the standard value of 180$^{\circ}$ usually expected if one
assumes that the accretion occurs at the magnetic pole itself.\\
In this varying geometry, we have fully computed the phasing of the radial 
velocity curve produced by material falling down the magnetic lines 
just above the white dwarf surface. 
The amplitude of the shift is determined as soon as $\theta_d$ and the 
beat period are fixed. 
The curve plotted in Figure 9.b) is computed for values of  
$\theta_d=12^{\circ}$ and a beat period of 14 days, and fits the data 
reasonably well.
The RV phase slowly varies with the beat period as observed, distorting the
period determination. 
The sudden change in the phasing when the pole switches, also well reproduces 
the magnitude of the phase jump ($\sim 50^{\circ}$) observed in the UV 
O-C measurements. It appears twice during a beat cycle. 
A beat value can be evaluated from the point distribution as $14.5\pm1.5$ days.
A pole-switching behaviour has been also suggested from photometric data by
Silber (1995, Fig.3). 
However large phase uncertainties are introduced, in this case, by the fact 
that the shapes of the light curves are variable and strongly depart from a 
sinusoidal curve. \\
Strictly speaking, one does not expect to observe the same shape of the beat
modulation for lines formed close to the white dwarf and for lines emitted 
at other positions along the accretion column. 
The emitting region is also moving, depending on the line production 
mechanism, and may introduce additional drifts. 
The present knowledge of the detailed emission line processes at work in 
these systems is not yet sufficient to allow a better evaluation of this 
effect. \\
An interesting consequence of this phase-drift model is the fact that any
period determination is biased depending on the length of the observation. 
Measurements extended on more than a beat period will reveal the orbital 
period, while data obtained in a few days will show either a shorter period 
than the orbital one or will not allow any period determination if situated 
close to the jump.  Thus a large spread of period values may result as it is 
indeed observed in Fig. 8. 
The considerations above also apply to the broad line components usually 
thought to be formed in the accretion column. 
Interestingly, the radial velocities of the \ha{} broad component  
measured by Sauter (quoted in Zucker et al. 1995) also show variations
at the orbital period ($\Omega$ frequency), while they have been found 
to be modulated with the short spin period ($\omega$ frequency)
by SBIOR, based on a set of data spread over six nights only. 
We predict that the O-C measurements by Sauter would mimic the same behaviour 
as for the NV line.        \\
In addition our reanalysis of 
Piirola data has  shown that a long period is also present in the 
polarization flux (see Section 5.3), together with an indication of the
(2$\omega-\Omega$) combination period. These periods are indeed expected 
for a cyclotron emission produced at the basis of the accretion column 
(Wynn \& King, 1992). 
By combining the two periods of 3.3308h and 3.3749h found in
Piirola polarization data,  a short (2$\omega-\Omega$)  period of 3.2878h
is derived independently, quite consistent with values determined by 
Silber (1995) and Mason et al. (1995a,b) from photometric data.
From the same two values, a long beat period ($\omega-\Omega$) of 
10.621$\pm$3$~10^{-3}~$days is also derived.
This value is not strictly consistent with the range  of values (13-16 days)
determined from the UV data, and with the similar 14 day period  suggested 
by Mason et al. (1995a) and Silber (1995).\\ 

In conclusion, we have shown that the period determinations are biased
depending on the temporal extension of the set of data. 
The phase-drift model described above  explains the inability to determine a 
unique adequate value for the  periods, when the white dwarf is not exactly 
synchronized. 
This simple picture has to be modified if one takes into account more 
physical complex configurations such as a multipole geometry 
(Mason et al. 1995), a decentered dipole, field line distortions at the 
threading region (Hameury, King \& Lasota 1986)  or possible inhomogeneous  
blobs in the infalling material. 
Moreover it is most probable that accretion would occur on both sides for 
intermediate configurations.    
The orbital period value is still inaccurate. It can, in principle, 
be unambigously established from the study of the absorption lines 
associated with the companion atmosphere.
A search for the Na lines was tentatively done but without success, implying 
that either the companion is very faint or that it is of an earlier spectral 
type than had been expected (Zucker et al. 1995). 
Finally, the discussion of the UV line formation suffers from the absence of 
a \gam-velocity value determination, which combined with the RV
amplitude should allow to constrain the emitting region in the accretion 
column. 
This can be solved  with the Hubble Space Telescope and a higher spectral 
resolution than provided by the IUE satellite.   \\

{\it Acknowledgments :}
We are grateful to Didier Pelat for providing us his program SPECTRE and to 
Paul Mason for useful discussions.
%________________________________________ Do not leave a blank line here!
%_____________________________________________________________________

\end{document}